\pdfoutput=1
\documentclass[RNAAS]{aastex631}

\shorttitle{Mega-Constellations and Astronomy}
\shortauthors{Petchetti}

\usepackage{etoolbox}
\makeatletter
\patchcmd{\thebibliography}{\parskip=1pt plus 1pt}{\parskip=1pt}{}{}
\patchcmd{\thebibliography}{\baselineskip=13pt plus 1pt}{\baselineskip=13pt}{}{}
\patchcmd{\thebibliography}{\baselineskip=13pt plus 1pt}{\baselineskip=13pt}{}{}
\patchcmd{\thebibliography}{\leftmargin=1.0em}{\leftmargin=0pt\leftskip=1.0em}{}{}
\patchcmd{\frontmatter@title@above}{\noindent{\sc Draft version \today}\\[2pt]}{\noindent}{}{}
\makeatother
\begin{document}

\title{Can Satellite Mega-Constellations Justify Their Impact on Astronomy?}

\author{Sai Charan Petchetti}
\correspondingauthor{Sai Charan Petchetti}
\email{saicharan@astronomy.org.in}

\collaboration{1}{}
\maketitle

In their latest paper, \citet{Rawls2020} drew attention to the impacts of satellite constellations on astronomy. Near-earth space is being increasingly commercialized by private space companies. This has many consequences for science, particularly, astronomy. Some estimates show that more than 100,000 satellites may orbit the Earth by 2030. This is an exponential increase from the approximately 4,000 operational satellites in orbit today. Satellite mega-constellations for satellite internet connectivity are one of the main drivers behind the explosion in the number of satellites. The largest of these private space companies alone is hoping to send more than 42,000 satellites into orbit.

Many researchers have raised concerns about the impact that these satellites may have on observational astronomy. The situation is particularly bad for very-wide-field imaging observations performed with large telescopes. In a recent study, the researchers estimated that approximately 30\% of the exposures of very-wide-field imaging observations performed with large telescopes can be ruined at the beginning and end of the night \citep{Hainaut2020}. An independent study published by the \textit{Rubin} observatory indicated a 40\% impact during the twilight observing time \citep{Rubin2020}.

The researchers discovered that wide-field imaging and spectroscopic surveys will be affected at a much lower level; up to 15\% of the exposures recorded during the beginning and end of the night will be ruined. Moreover, up to 0.5\% of medium-duration exposures with traditional fields of view will be ruined during twilight. By contrast, brief telescopic observations will largely remain unaffected.

The two most affected types of observations (\textit{i.e.}, very-wide-field and wide-field imaging) are the backbone of astronomy and the main tools for discovery. Hence, the adverse impact that satellite mega-constellations can have on key surveys is deeply worrying.

Proponents of satellite mega-constellations often cite the connectivity benefits these systems could offer in order to justify the impact on astronomy. There is no denying that affordable internet access is strongly needed. However, we must ask ourselves if these satellite mega-constellations will be able to meet the demand. The estimated set-up and maintenance costs of satellite internet connectivity are prohibitively expensive for all countries in greatest need of internet access. A recently published dataset \citep{RawlsData2020} has shown that the population that needs and can afford satellite internet connectivity is very small.

Considering the low utility of satellite internet connectivity to those who need it most, the current proposals of tens of thousands of satellites orbiting the Earth do not justify the impact they will have on the night sky and astronomy. The choice between protecting our night sky and satellite internet connectivity is not a binary one. Considering the current circumstances, it is impractical to call for the complete abandonment of all satellite mega-constellation projects. Greatly reducing the proposed number of satellites and regulating satellites more tightly would be more feasible.

\bibliography{references}{}
\bibliographystyle{aasjournal}

\end{document}